\documentclass[12pt]{article}
\usepackage[dvips]{graphicx}

\begin{document}

\author{W. Wen$^{1,3}$, P. Chau Huu-Tai$^{1}$, D. Lacroix$^{1}$, 
Ph. Chomaz$^{1}$ \\
and \\
S. Ayik$^{2}$ \\
1)GANIL, BP 5076, 14076 CAEN Cedex 5, France; \\
2)Tennessee Technological University, \\
Cookeville, TN 38505, USA  \\
3)Institute of Modern Physics, \\
Lanzhou 730000, P. R. China }
\title{Quantum and statistical fluctuations in dynamical symmetry breaking 
\footnote{This work is supported in part by the US DOE grant no.
 DE-FG05-89ER40530}}
\date{}
\maketitle

\begin{abstract}
Dynamical symmetry breaking in an expanding nuclear system is investigated
in semi-classical and quantum framework by employing a collective transport 
model  which is constructed  to mimic the collective behavior of expanding
systems.
It is shown that the fluctuations in collective coordinates during the 
expansion 
are developed mainly by the enhancement of the initial 
fluctuations by the driving force, and  that statistical and quantum 
fluctuations have similar consequences. 
It is pointed out that the quantal fluctuations may play an important role 
in the 
development of instabilities by reducing the time needed to break the 
symmetry, and the
possible role of quantal fluctuations in spinodal decomposition of nuclei is 
discussed. 
\end{abstract}



\section{Introduction}

\smallskip 
Dynamical symmetry breaking is a general behavior in nature and
is observed in many field of physics. For example, phase transitions are
often related to a dynamical symmetry breaking. The liquid-gas phase transition 
in infinite matter may be viewed as breaking of translational invariance which is 
related to mechanical instability of the system against small density fluctuations inside the spinodal zone. 
The spinodal  instability may provide a possible mechanism for
triggering the fragmentation of a hot piece of nuclear matter produced in
heavy-ion collisions \cite{Cugnon, BertschSiemens}. The composite system formed in the overlap
zone of the colliding ions expands and cools down. 
Depending on the initial compression and temperature,
it may enter the unstable region of
the phase diagram. In such a scenario, the fragment formation takes place by
a rapid growth of density fluctuations in the spinodal region. Such a scenario is  valid 
for infinite matter or a  finite system. In matter, unstable modes are sound modes characterized 
by plane waves, whereas in a finite system these are collective
vibrations associated with density fluctuations with multipolarity $L$ and $M$. For example,
a spherical system can break into pieces as soon as the collective mode with 
multipolarity  $L$  larger than one becomes unstable. 

Stochastic one-body transport models provides a suitable framework for a theoretical 
description of the collision process all the way from the initial stage up to the final state 
involving development of large density fluctuations and formation of fragments 
\cite{AyikGregoire, RandruprRemaud,Abe}. This approach in a semi-classical approximation for the phase-space density, 
which is frequently referred as the Boltzmann-Langevin model, 
has been applied to investigate the spinodal decomposition of nuclear systems 
\cite{Alfio}. Stochastic Time Dependent Hartree Fock (STDHF) 
approaches are now under development in order to take into account the quantal 
nature of single particle motion \cite{Balian,Suraud}. 

A valuable insight on dynamical symmetry breaking, and the role of quantal and statistical fluctuations on 
the instabilities can be gained by investigating the early development of density fluctuations. The linear 
response treatment of the one-body stochastic transport  models, in semi-classical or quantal framework,  
provides a useful basis for this purpose. In a recent work \cite{Jacquot}, the dominant collective 
modes in an expanding finite nuclear system are identified in a linear response treatment by solving  
temperature dependent quantal RPA equation, and the properties of these collective modes are investigated 
as a function of dilution of the system. These calculations illustrate that, as the system expands, the collective 
modes which were all stable at normal density, become  softer until their energies become zero and then 
bifurcate toward the imaginary values, signaling the instability of the corresponding mode. 

The density 
fluctuations associated with the collective degrees of freedom can be determined in terms of collective transport 
equations. These transport equations for the collective variables can be deduced from the microscopic 
transport models  by projecting the equation of motion onto a specific collective 
mode \cite{Ayik, Norenberg}, or can be derived by 
some other means \cite{Hofmann}. Since, the standard one-body transport 
approach provides a classical description of  
the collective motion , the deduced collective models are best suited at high excitation energies when the 
temperature is much larger than the typical frequencies of the collective motion. On the other hand, when 
the characteristic energies of the collective modes are comparable with or larger than temperature of the system, 
the quantal fluctuations of the collective dynamics become important or even dominant. Then, a quantum 
transport approach of the collective variables is required for describing dynamics of large density 
fluctuations. The stochastic one-body transport description can be improved by incorporating the memory 
effects associated with finite duration of binary collisions. 
As demonstrated in a recent work \cite{AyikZ,AyikRandrup}, in such non-Markovian description, transport properties of collective motion evolve in accordance with the quantal 
fluctuation-dissipation relation \cite{Gardiner}.

In this work, we consider a single collective mode, specifically the quadru-
pole moment of the fluctuating 
density in an expanding nuclear system and study its evolution by employing collective transport models, in 
both classical and quantal forms. Although these models can be derived from the underlying microscopic 
transport models in Markovian and non-Markovian forms, we follow a phenomenological approach and 
parameterize the basic ingredients of the model in accordance with the recent TDHF calculations for 
expansion of a hot spherical system  \cite{Vautherin,Lacroix}, and the RPA calculations for 
describing the unstable collective modes \cite{Jacquot}.  In the 
Fokker-Planck approach, the coupling between the collective mode and the intrinsic degrees of freedom is 
described in terms of  friction and diffusion coefficients. This simple but realistic model allows us to calculate 
the development of fluctuations associated with the collective mode during the expansion of the nuclear 
system, and to investigate the role of quantal and statistical 
fluctuations in the dynamics of the 
symmetry breaking.

\section{The model}

\subsection{The collective Hamiltonian}

Combining the results of the finite temperature TDHF description of the
expansion of hot nuclear sources ref. \cite{Vautherin,Lacroix} and the
RPA investigation of the stability of a finite hot expanding nuclear source 
\cite{Jacquot}, we can determine the most important collective modes
and describe their transitions from stable to unstable configurations as a
function of expansion, temperature and source size. Here, we specifically
consider evolution of the quadrupole shape fluctuations in an expanding
spherical hot nuclear source. According to finite temperature TDHF
calculations with initial temperatures in the range of $T=5-10$ MeV,
the system executes monopole vibrations and cools down by particle
evaporation.  Depending on the size of the source and the initial temperature, it takes between $30$ fm/c and $75$ fm/c to reach the maximum dilution that occurs typically around  $1/3$ of the normal nuclear matter density. 
According to the constraint RPA calculations, the quadrupole mode with an initial frequency $\omega $, 
which is found to be about $\omega \simeq 65A^{-1/3}$ MeV $\hbar$ at moderate temperature, 
becomes softer as the system expands. It eventually becomes unstable and reaches an imaginary frequency 
of about $i20A^{-1/3}$
MeV/$\hbar $ at the maximum dilution. 

In order to describe the evolution of the quadrupole mode during the
expansion of the source, we consider the following model, in which the
Hamiltonian of the collective motion is taken as 
\begin{equation}
H=\frac{P^{2}}{2M}+U\left( Q\right)
\end{equation}
where $M=3AmR_{0}/8\pi $ is the irrotational mass parameter associated with
the collective variable $Q$ determined by the sharp nuclear surface $R_{0}$,
$A$ is the number of nucleons in the source and $U(Q)$ represents the
collective potential energy. For small deformation $Q$ , the potential energy has a harmonic form $ U\left( Q\right) = K\left( t\right)  Q^{2}/2+...$ with a time dependent spring constant $ K\left( t\right) = K_{0}g(t)$ 
where $K_{0}=\omega^{2}M$ and 
$\omega =65A^{-1/3}$ MeV/$\hbar $ denotes
the frequency of the giant quadrupole vibrations at normal density. The
time dependent factor $g(t)$ in the spring constant describes the transition from stable to
unstable regime during the expansion of the source. This factor is parameterized In accordance with the TDHF and RPA calculation: starting from $g(0)=1$ it monotonicaly decreases in time
and reaches a value around  $g(t_{0})=-1/9$ at a characteristic time $t_{0}$ for the source to reach the maximum instability (typically, $t_{0}= $ 50 fm/c).  
The harmonic approximation is valid during early stages of the unstable evolution when the magnitude of fluctuations is small. In order to describe large fluctuations, we need to incorporate an anharmonic terms in the potential energy,
$U\left( Q\right) $, which prevent the fluctuation to grow indefinitely without
bound. For symmetry considerations,  we add an anharmonic term of the form 
$Q^{4}$ in the potential energy to achieve the saturation of the fluctuations
\begin{equation}
U\left( Q\right) =\frac{K\left( t\right) }{2}Q^{2}+\beta Q^{4}
\end{equation}
Since, we are mainly concern with  the early development of 
fluctuations, the precise form of this non-linear term is not  important for this
purpose. The numerical coefficient in front of the non-linear
term is taken to be $\beta=K_{0}/2$ in order to obtain saturation of fluctuations
at a reasonable value of the collective variable $Q$ which corresponds to the 
quadrupole deformation of  the
fragmenting system (e.g. the quadrupole deformation may lead to break-up 
of a system into two pieces). The numerical calculations presented are performed
for a system containing $A=64$ nucleons.

\subsection{Classical Langevin dynamics}

The classical Langevin approach provides a useful framework for describing
the dynamics of the collective motion including dissipation and
fluctuations. In this approach, the trajectory in the collective phase
space, $\left( Q^{\left( i\right) }(t),P^{\left( i\right) }(t)\right) $ of
an event $i$, is determined by solving a Langevin equation, 
\begin{eqnarray}
\frac {dQ^{\left( i\right) }}{dt} &=&\frac{P^{\left( i\right) }}{M} \\
\frac {dP^{\left( i\right) }}{dt} &=&\frac{\partial U\left( Q^{\left( i\right)
}\right) }{\partial Q}-\gamma P^{\left( i\right) }+\xi ^{\left( i\right)
}\left( t\right) 
\end{eqnarray}
where $\gamma $ is the friction coefficient and $\xi ^{(i)}(t)$ is the random
force arising from the coupling of the collective degrees of freedom with the
single particle motion. The fluctuating force $\xi ^{(i)}(t)$ is assumed to
be a white noise, and specified by a Gaussian distribution with zero mean 
$<\xi (t)>=0$ and a second moment 
\begin{equation}
\left\langle \xi \left( t\right) \xi \left( t^{\prime }\right) \right\rangle
=\delta \left( t-t^{\prime }\right) 2D
\end{equation}
where $\left\langle ...\right\rangle $ represents the statistical average
over the ensemble of events $i.$ The diffusion coefficient $D$ is determined in
terms of the friction coefficient by the classical dissipation-fluctuation
relation 
\begin{equation}
D=\gamma MT  \label{Einstein}
\end{equation}
where $T$ represents the time-dependent temperature of the internal system
which may be specified by assuming an isentropic expansion of the nuclear
source. The only remaining parameter of the model is the friction
coefficient, which may be estimated from the observed damping width of the
giant quadrupole excitation according to $\gamma =\Gamma /\hbar $. In the 
calculations, we take the damping width to be $\Gamma=85 A^{-2/3}$ MeV.

By generating a sufficiently large number of events, we can construct the
phase-space density $f(P,Q,t)$ associated with the collective variables $P$
and $Q$ as 
\begin{equation}
f\left( Q,P,t\right) =\frac{1}{N}\sum_{i=1}^{N}\delta \left( Q-Q^{\left(
i\right) }(t)\right) \delta \left( P-P^{\left( i\right) }(t)\right) 
\end{equation}
where $N$ denotes the number of events. The probability distribution $n(Q,t)$ of $%
Q$ and the variances of the collective variables $\sigma _{Q}(t)$ and $%
\sigma _{P}(t)$ are calculated as 
\begin{equation}
n\left( Q,t\right) =\int dP\;f\left( Q,P,t\right) =\frac{1}{N}%
\sum_{i=1}^{N}\delta \left( Q-Q^{\left( \lambda \right) }(t)\right) 
\end{equation}
and 
\begin{eqnarray}
\sigma _{Q}(t) &=&<Q^{2}>=\int dPdQ\;Q^{2}f\left( Q,P,t\right)  \\
\sigma _{P}(t) &=&<P^{2}>=\int dPdQ\;P^{2}f\left( Q,P,t\right) 
\end{eqnarray}
The initial conditions of the events generated by the Langevin equation is
specified by the Boltzmann distribution with the initial temperature $T_{0}$
\begin{equation}
f_{0}\left( Q,P\right) =\frac{1}{Z}\exp \left( -\frac{H(Q,P)}{T_{0}}\right). 
\end{equation}
In our simulations, we sample the initial fluctuations using a metropolis
algorithm\cite{Metropolis}.

The classical description of the Langevin equation provides a good approximation 
at sufficiently high temperatures at which the dynamics is dominated
by thermal fluctuations. At low temperatures, the Langevin approach may 
be improved by incorporating the quantal fluctuations through the initial 
conditions and by specifying the diffusion coefficient through the quantal
fluctuation-dissipation relation \cite{AyikZ,AyikRandrup}. As a matter of fact, 
this description produces exact
quantal phase-space density in the harmonic limit, for which the quantal
fluctuation-dissipation relation may be expressed as $D=(\gamma M\hbar
\omega /2)\coth (\hbar \omega /2T)$ for real $\omega $. However, in general for 
non-linear evolution, collective motion should be investigated in the
basis of a quantal transport equation.

\subsection{Quantal transport model}

In quantal approaches the system is described by its density matrix $\hat{%
\rho }(t)$ on the collective Hilbert space, $\hat{\rho}(Q,Q^{\prime
},t)=\left\langle Q\right| \hat{\rho}(t)\left| Q^{\prime }\right\rangle ,$
which is determined by a quantal transport equation 
\begin{equation}
i\hbar \frac{\partial }{\partial t}\hat{\rho}(t)=\left[ \hat{H}(t),\hat{\rho}%
(t)\right] +i\hbar \hat{K}(\hat{\rho})
\end{equation}
where $K(\hat{\rho})$ represents a ''collision term'' arising from the
coupling of the collective motion with the intrinsic variables, which
corresponds to the friction and diffusion term in the Langevin equation. The
collective density matrix may be expanded in terms of a complete set of
time-dependent wave functions $|\Psi _{i}(t)>$ as 
\begin{equation}
\hat{\rho}(t)=\sum_{i,j}\left| \Psi _{i}(t)\left\rangle \rho
_{ij}(t)\right\langle \Psi _{j}(t)\right| 
\end{equation}
where the time-dependent wave functions are obtained by solving the
time-dependent
Shr\"{o}dinger equation 
\begin{equation}
i\hbar \frac{\partial }{\partial t}\left| \Psi _{i}(t)\right\rangle =\hat{H}%
(t)\left| \Psi _{i}(t)\right\rangle 
\end{equation}
and the elements of  density matrix $\rho _{ij}(t)$ obey  a master equation 
\begin{equation}
\frac{\partial }{\partial t}\rho _{ij}(t)=<\Psi _{i}(t)|K(\hat{\rho})|\Psi
_{j}(t)>
\end{equation}
If  the system  initially is in statistical equilibrium at an initial temperature $T_{0}$,
it can be represented by a density matrix, 
\begin{equation}
\hat{\rho}\left( t=0\right) =\frac{1}{Z}\exp \left( -\frac{\hat{H}_{0}}{T_{0}}\right)
\end{equation}
It is convenient to introduce the eigenstates $|\varphi _{i}>$ and
the eigen-energies $E_{i}$ of the initial Hamiltonian $H_{0}=H(t=0)$%
\begin{equation}
H_{0}|\varphi _{i}>\;=\;E_{i}|\varphi _{i}>
\end{equation}
as initial basis. 
Consequently, the initial density matrix reduces to Boltzmann occupation
probabilities of the eigenstates of the initial Hamiltonian, 
\begin{equation}
\rho _{ij}\left( t\right) =\delta _{ij}\frac{1}{Z}\exp \left(-\frac{E_{i}}{T_{0}}\right).
\end{equation}

The probability density $n(Q,t)$, and the variances $\sigma _{Q}(t)$ and $%
\sigma _{P}(t)$ of the collective variables are calculated according to 
\begin{equation}
n(Q,t)=\left\langle Q\right| \hat{\rho}(t)\left| Q\right\rangle
\end{equation}
and 
\begin{equation}
\sigma _{Q}(t)=<\hat{Q}^{2}>={\rm tr}\hat{Q}^{2}\hat{\rho}(t)
\end{equation}
\begin{equation}
\sigma _{P}(t)=<\hat{P}^{2}>={\rm tr}\hat{P}^{2}\hat{\rho}(t)
\end{equation}
Since the model Hamiltonian is nearly harmonic at the initial time $t=0$,
the initial variances of the collective variables can approximately be
expressed in a closed form as 
\begin{equation}
\begin{array}{l}
\sigma _{Q}(0)=\left \langle Q^{2}\right \rangle _{0}\simeq \left( \hbar \omega/ 2K_{0}\right) 
\coth \left( \hbar \omega/ 2T_{0}\right) 
\begin{array}{l}
\\ 
\overrightarrow{T_{0}\rightarrow 0}
\end{array}
\hbar \omega / 2K_{0} \\ 
\sigma _{P}(0)=\left\langle P^{2}\right\rangle _{0}\simeq \left( M\hbar
\omega /2 \right) \coth \left( \hbar \omega / 2T_{0}\right) 
\begin{array}{l}
\\ 
\overrightarrow{T_{0}\rightarrow 0}
\end{array}
M\hbar \omega / 2
\end{array}
\label{fluct_quant}
\end{equation}
Here, the approximate expressions are valid when temperature is small as compared to
the collectives frequency, $T_{0}\ll \hbar \omega $, and hence the initial
variances are dominated by the ground state zero point fluctuations. 
In the opposite limit when temperature is much larger than 
half the initial collective energy  $\hbar\omega \ll 2T$
the initial variances of the collective variables can be approximated by 
\begin{equation}
\begin{array}{l}
\sigma _{Q}(0)\simeq T_{0}/ K_{0} \\ 
\sigma _{P}(0)\simeq MT_{0}
\end{array}
\label{fluct_class}
\end{equation}
which corresponds to the classical description discussed in the previous section. 

\smallskip

\section{Results }

We carry out a number of simulations on the basis of two different
approaches presented above to investigate the development of quadrupole 
shape instabilities in an expanding spherical nuclear system prepared at
a range of initial temperatures and calculate the probability density $n(Q,t)$
and the variances $\sigma _{Q}(t)$, $\sigma _{P}(t)$ of the collective
variables. In our calculations, we consider that the initial state is prepared at normal
density and at a temperature in the range of  $T=5-10$ MeV. This is in accordance with the 
BUU 
calculations for central heavy-ion collisions around Fermi energy, which
show that the collisions lead to an equilibrated composite spherical system 
at temperatures around $T=10$ MeV. The second important parameter
 in our model is the characteristic time $t_{0}$ for expansion until the system reaches the turning point, which has a typical magnitude around 
$t_{0}=50$ fm/c. In order to see the influence of the expansion time on the dynamics of symmetry
breaking, we present calculations for different $t_{0}$.

\smallskip

\subsection{Classical dynamics}

\begin{figure}[tbph]
\begin{center}
\includegraphics*[height=12cm,width=15cm]{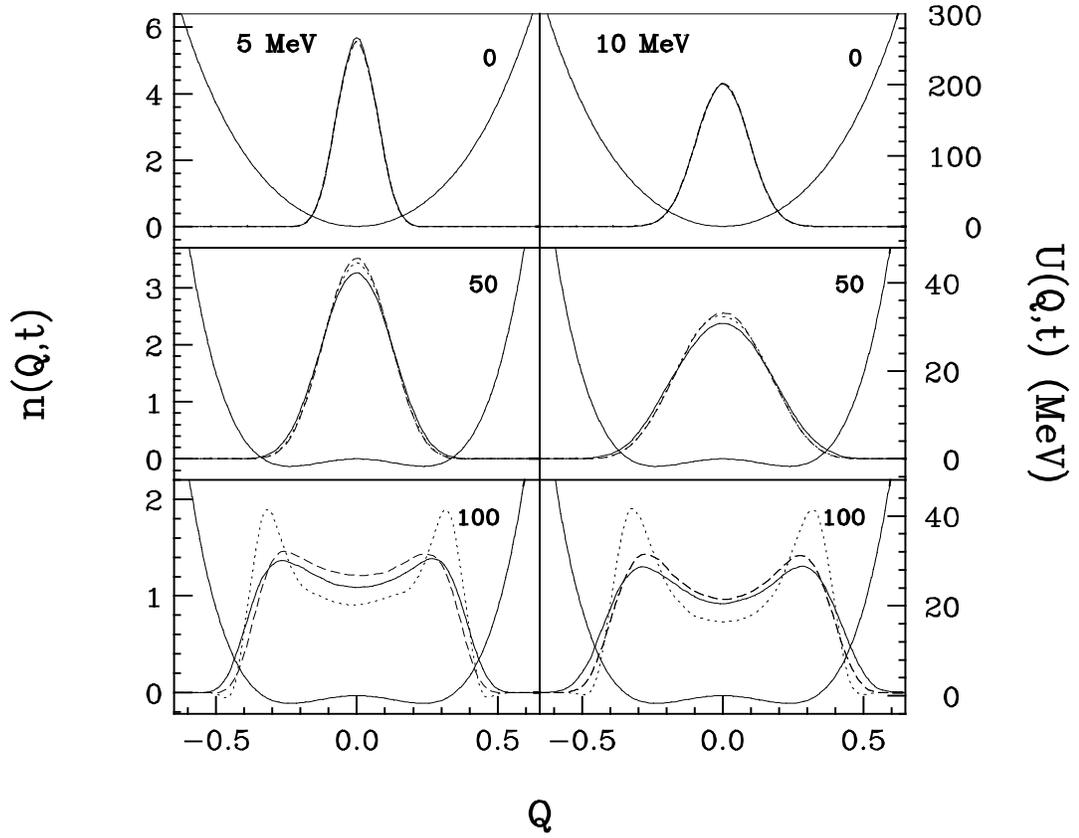}
\end{center}
\caption{ Probability distribution of collective variable and potential energy as a function of $Q$
at times $t=0, 50, 100$ fm/c for initial temperatures $T_{0}=5$ MeV(left panel) and $T_{0}=10$ MeV
(right panel). Solid, dashed and dotted lines are the Langevin simulations with constant initial temperature, the Langevin simulation with time dependent temperature and the pure potential 
calculations, respectively. }
\label{fig1}
\end{figure}

The result of simulations of the classical Langevin equation is illustrated in 
figure 1 by solid lines, in which 
the probability distribution $n(Q,t)$ and the potential energy $U(Q,t)$ 
are plotted as a function of $Q$ at 
several times for two different initial temperatures $T_{0}=5$ MeV (left panel) and $T_{0}=10$ 
MeV (right panel).
In these simulations temperature is assumed to be constant and equal to the initial value. The width of 
distribution rapidly increase as the potential gets softer and  the probability distribution splits into two 
components after the potential exhibits a local maximum at $Q=0$. Figure 2  displays the variances associated with 
the collective variable $Q$ and the collective momentum $P$ as a function of time for the initial 
temperatures $T=5$ MeV and $T=10$ MeV. The fluctuation in the collective coordinate rapidly increase 
until the system reaches the maximum instability and then saturates after 100 fm/c. During the same time the 
momentum distribution remains almost constant and goes to the asymptotic value given by the equipartition 
theorem $\sigma _{P}\simeq MT_{0}$. In figures 1 and 2,
the dotted lines display the results of the pure potential calculations without the dissipation and fluctuations. 
During the expansion phase until the source reaches maximum dilution at about $t=50-60$ fm/c, the pure potential result for $n(Q,t)$ is very close to the Langevin simulations with friction term and stochastic force. Therefore, the fluctuations of 
the collective variable $Q$ during the expansion phase are not developed dynamically 
but originate mainly from the propagation of the initial statistical fluctuations by the driving force. As the system expands, it  cools down and consequently 
the magnitude of  the stochastic force in the 
Langevin equation become smaller. The dashed lines in figures 1 and 2
indicate the simulations carried out by employing a time dependent temperature, 
which is determined according to 
$T \left( t \right) =T_{0}\left (R_{0}/ R \left (t \right) \right) ^2 $
by assuming adiabatic expansion, where $ R \left (t \right)$ represents the 
root-mean-square radius of the expanding system.
As seen from these, the cooling has a minor effect on the probability distribution 
$n(Q,t)$ during the early phase of the expansion. However,  cooling has a
sizable effect in momentum space at large times, as  
seen in the asymptotic value of the $ \sigma _{P}$ 
which also come close to the result obtained by the pure potential calculations.

\begin{figure}[tbph]
\begin{center}
\includegraphics*[height=10cm,width=13cm]{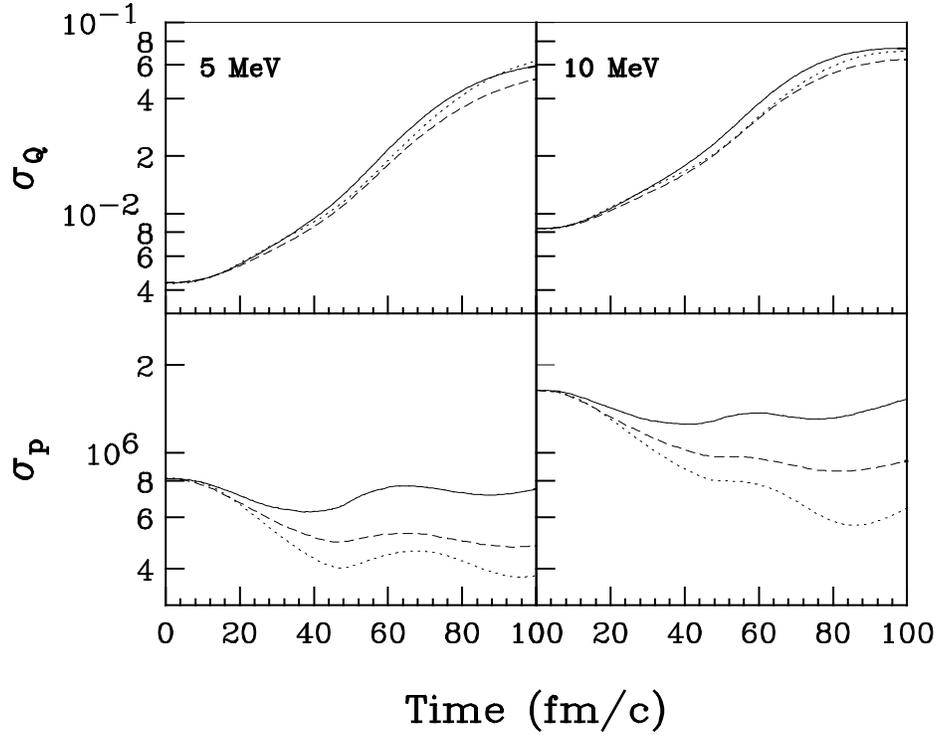}
\end{center}
\caption{Variances associated with collective variable $Q$ and collective momentum $P$ as a function
of time at initial temperatures $T_{0}=5$ MeV(left panel) and $T_{0}=10$ MeV
(right panel). Solid, dashed and dotted lines are the Langevin simulations with constant initial temperature, the Langevin simulation with time dependent temperature and the pure potential 
calculations, respectively.}
\label{fig2}
\end{figure}

\smallskip

\subsection{Comparison between quantum and classical dynamics}

\begin{figure}[tbph]
\begin{center}
\includegraphics*[height=12cm,width=15cm]{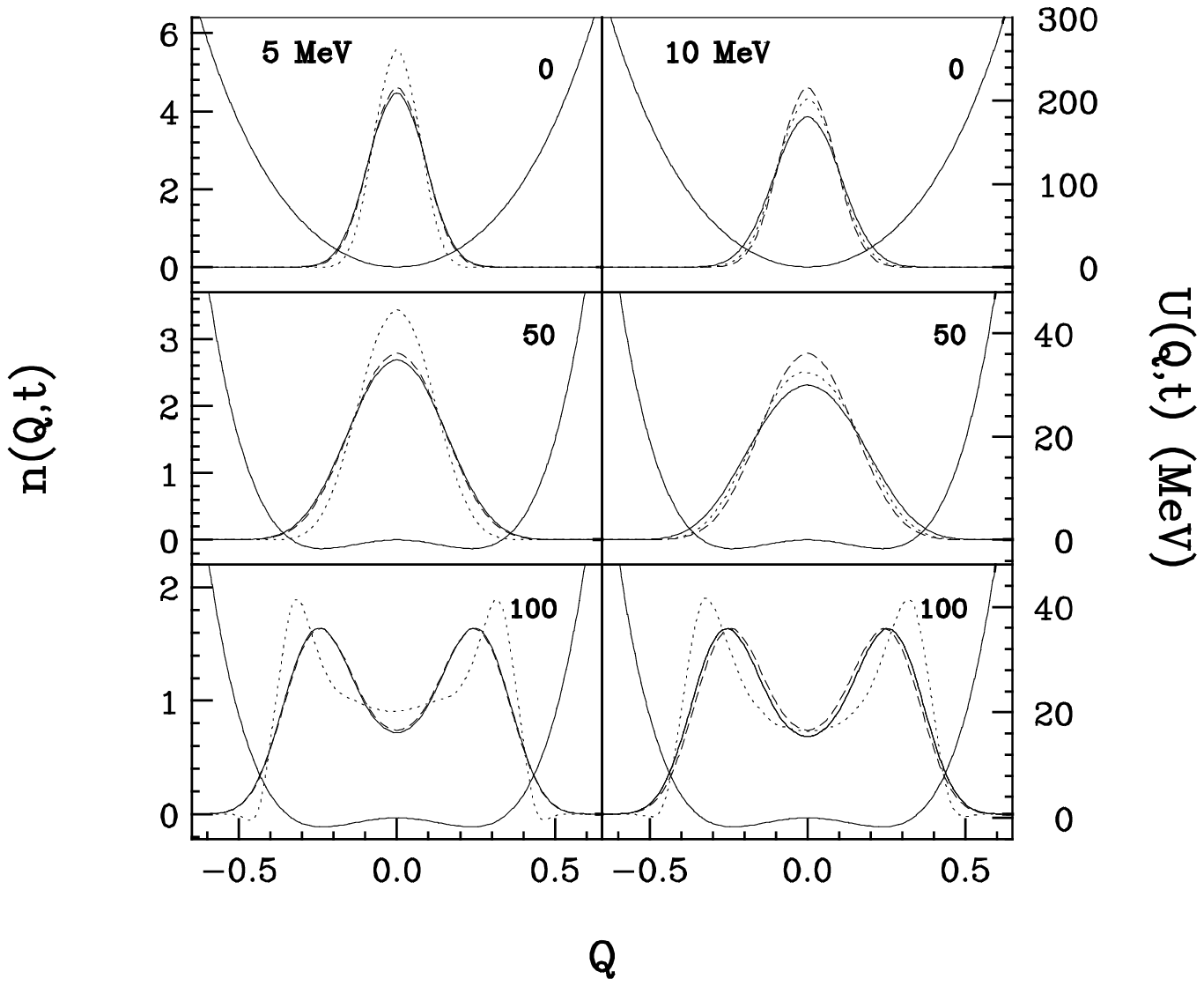}
\end{center}
\caption{ Probability distribution of collective variable and potential energy as a function of $Q$
at times $t=0, 50, 100$ fm/c for initial temperatures $T_{0}=5$ MeV(left panel) and $T_{0}=10$ MeV
(right panel). Solid, dashed and dotted lines are the full quantal calculation, the quantal calculations with ground state, and the classical calculations without friction and stochastic forces, respectively.}
\label{fig3}
\end{figure}

According to the classical Langevin simulations, during the expansion phase, 
the fluctuation of collective variable $Q$ is mainly determined by the 
potential evolution and the friction and 
stochastic forces have a minor effect. It is 
reasonable to expect  a similar behavior in
the quantal evolution of the collective motion. 
Therefore in the application of the quantal model presented 
in section 2.3,  we neglect the collision term, and calculate 
the probability distribution $n(Q,t)$ and the variances $\sigma _{Q}(t)$, 
$\sigma _{P}(t)$ by keeping the occupation 
probabilities $\rho _{i}$ to be constant and equal 
to initial values specified by the Bolztmann factors. 
Solid lines in figure 3 illustrates the result of quantal 
calculations of the probability distribution $n(Q,t)$ 
at different times for two initial temperatures $T=5$ MeV (left panel) and 
$T=10$ MeV (right panel). In 
the same figure, dotted lines show the classical simulations without 
friction and stochastic force, and also, as a 
reference dashed lines show the quantal calculations performed 
only with the ground state (ie the $T=0$ 
MeV case). 
\begin{figure}[tbph]
\begin{center}
\includegraphics*[height=10cm,width=13cm]{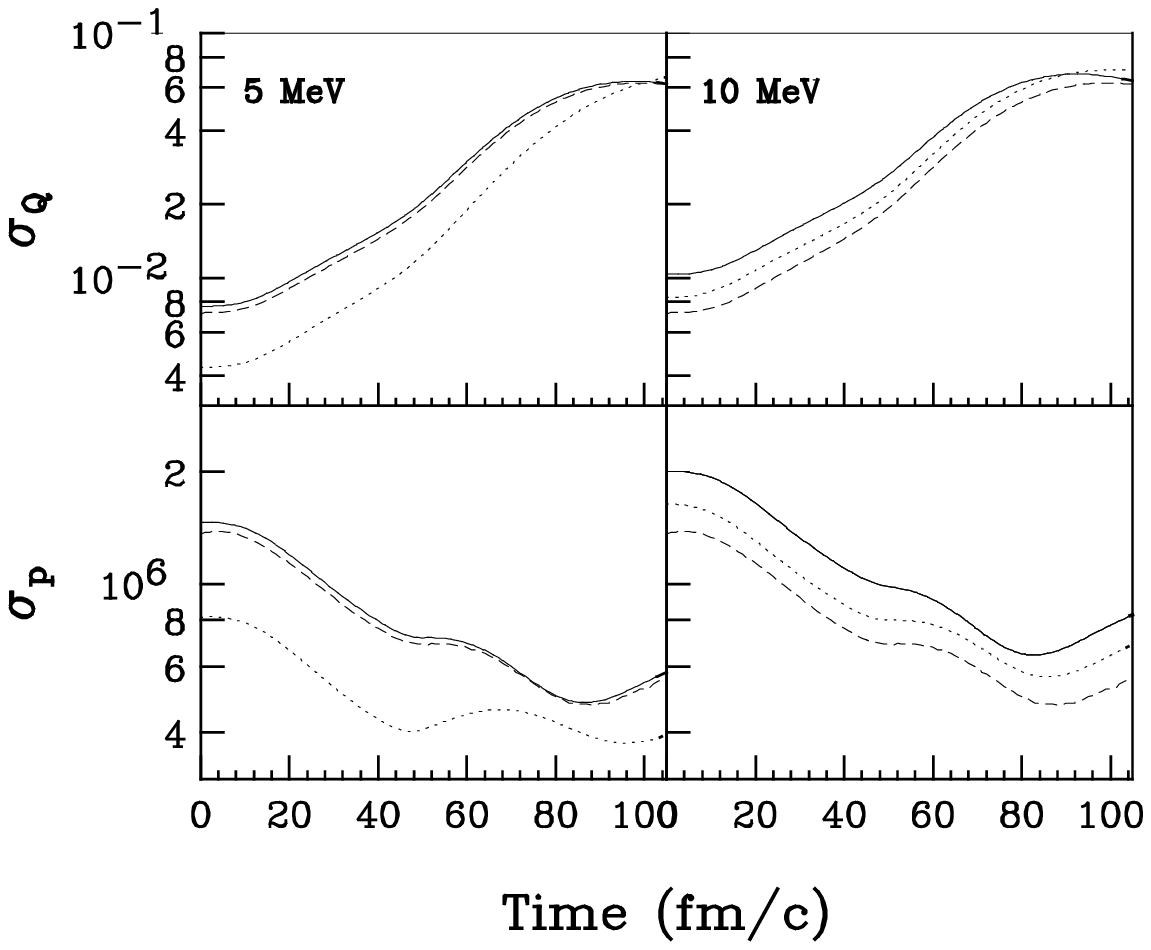}
\end{center}
\caption{Variances associated with collective variable $Q$ and collective momentum $P$ as a function
of time at initial temperatures $T_{0}=5$ MeV(left panel) and $T_{0}=10$ MeV
(right panel). Solid, dashed and dotted lines are the full quantal calculation, the quantal calculations with ground state, and the classical calculations without friction and stochastic forces, respectively. }
\label{fig4}
\end{figure}
Figure 4 illustrates the evolution of the variances 
$\sigma _{Q}(t)$ and $\sigma _{P}(t)$ as a function of time for 
initial temperatures $T=5$ MeV and $T=10$ MeV. The classical 
simulations, the full quantal calculations and the quantal 
calculations with ground state are shown by dotted, 
solid and dashed lines, respectively. The fluctuations of 
the collective variable, $\sigma _{Q}(t)$, exhibit 
an exponential growth as soon as the mode becomes unstable, 
and saturate at large times, in both the classical and 
quantal calculations. The exponential growth rates are 
the same in both calculations, but the 
magnitude of fluctuations are larger in the quantal case. 
This is illustrated in Figure 5 which shows the critical 
time $\tau$ it takes to
reach $\sigma_{Q}=0.2$ as a function of the expansion time $t_{0}$ of the
source for the initial temperatures $T=5$ MeV and $T=10$ MeV. As seen, the
quantal fluctuations shown by dashed lines reduce the time it takes to reach
a finite value of the fluctuations. The reduction of this time, depending on
the temperature of the system, can be as large as factor of two. 
For comparison, figure 6 shows different classical calculations 
of the critical time $\tau$ as a function of the 
expansion time $t_{0}$. In this figure, the Langevin simulations 
with constant initial
temperature, the Langevin simulations with time dependent 
temperature and the pure potential calculations are indicated by solid, 
dashed and dotted lines, respectively.

\begin{figure}[tbph]
\begin{center}
\includegraphics*[height=10cm,width=10cm]{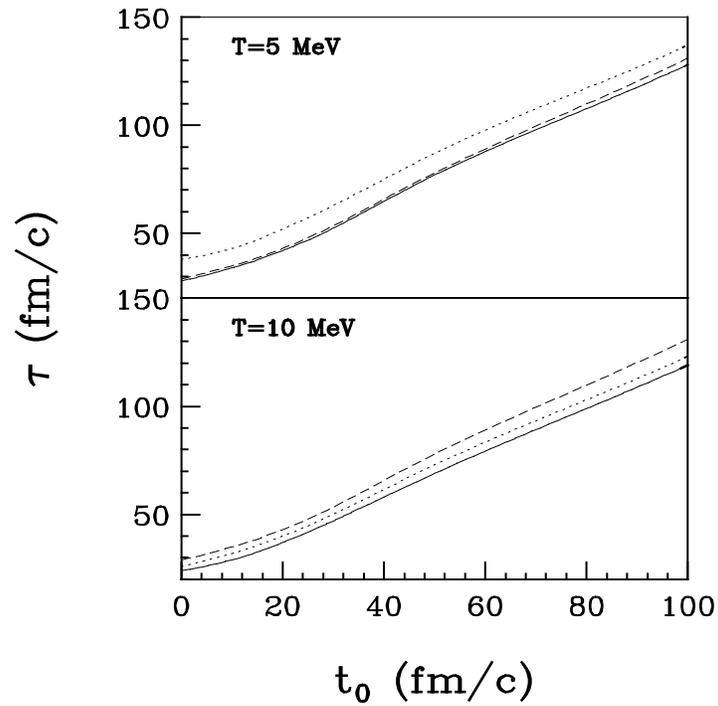}
\end{center}
\caption{Time $\tau$ that takes to reach $\sigma_{Q}=0.2$ as a function of the expansion
time $t_{0}$ at temperatures $T=5$ MeV (top panel) and $T=10$ MeV (bottom panel). Solid, dashed and dotted lines are the full quantal calculation, the quantal calculations with ground state, and the classical calculations without friction and stochastic forces, respectively.}
\label{fig5}
\end{figure}

\begin{figure}[tbph]
\begin{center}
\includegraphics*[height=10cm,width=10cm]{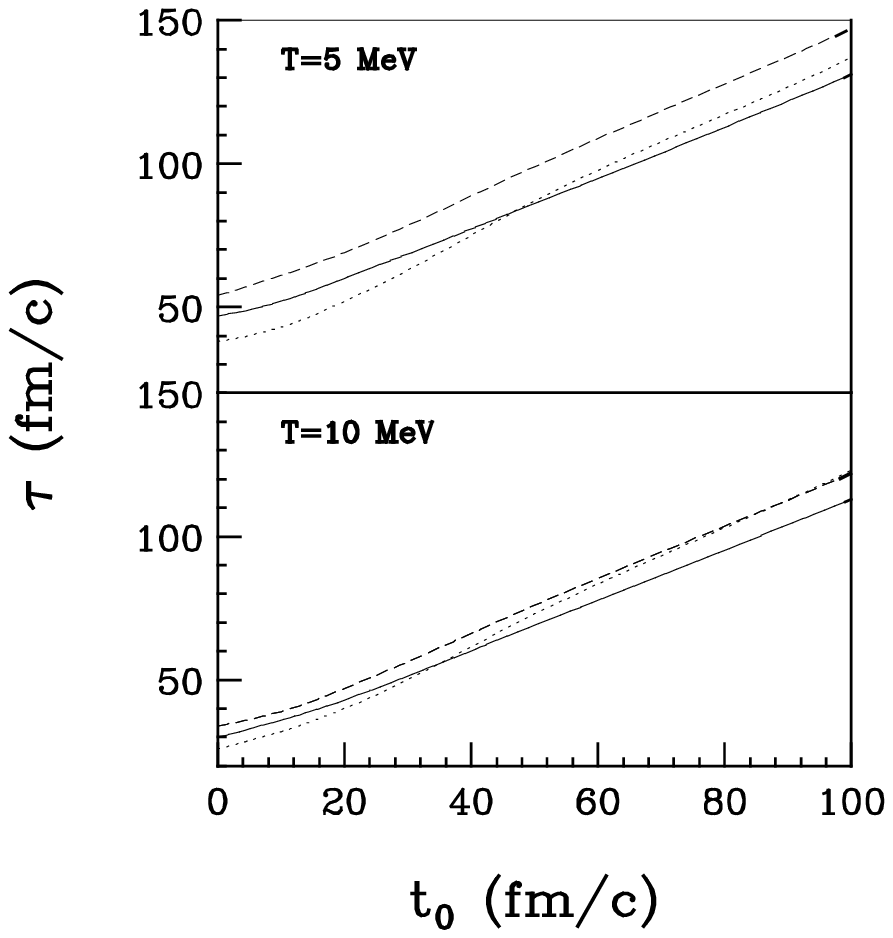}
\end{center}
\caption{Same as in figure 5. Solid, dashed and dotted lines are the Langevin simulations with constant initial temperature, the Langevin simulation with time dependent temperature and the pure potential 
calculations, respectively.}
\label{fig6}
\end{figure}

\smallskip

\section{Conclusions}
We investigate the development of density fluctuations associated with collective modes 
(here we specifically consider quadrupole mode) in an expanding nuclear system by employing 
phenomenological  classical and quantal transport equations.
According to mean-field calculations, a finite nuclear system 
prepared at moderate temperatures around $T=5-10$ MeV expands 
until the turning point which is located inside the unstable zone, 
and then continue to execute monopole vibrations. We calculate the evolution 
of the probability distribution of the collective coordinate $Q$ and the 
collective momentum $P$ and determine the variances associated with 
these distributions as a function of time.  Our investigations 
indicate that during the expansion phase, the dissipation and 
fluctuation mechanism play an important role in the evolution 
of the momentum distribution, but  the fluctuations in collective 
coordinate $Q$ are developed mainly by enhancement of the initial 
fluctuations of statistical or quantal origin. Furthermore, it 
appears that quantal fluctuations, which are already present at 
the initial state,  play an important role in particular at low temperatures,
 in dynamics of symmetry 
breaking by reducing the time required for reaching the critical 
fluctuations. Consequently, 
the spinodal decomposition and the resultant fragmentation  may take 
place in a faster time scale than predicted by semi-classical simulations 
\cite{Alfio,Marie}. Therefore,
it is off  great interest to develop stochastic 
simulation methods for heavy-ion collisions by incorporating 
quantal fluctuations associated with collective motion in a 
suitable manner. 

\smallskip

{\bf Acknowledgments}

Two of us (S. A. and W. W.) gratefully acknowledges GANIL Laboratory for a partial support and warm hospitality extended to them during their visit to Caen. Two of us (Ph. C. and D. L.) thank the Tennessee Technological University for a partial support and warm hospitality during their visit. Also, W. W. 
gratefully acknowledges Chinese Academy of Sciences for financial support.
This work is supported in part by the US DOE grant no. DE-FG05-89ER40530.


\smallskip

\begin{thebibliography} {99}
\bibitem{Cugnon} 
{J. Cugnon, Phys. Lett. {\bf 135B} (1986) 374.}
\bibitem{BertschSiemens} 
{G. Bertsch and P.J. Siemens, Phys. Lett. {\bf 126B}
(1983) 9.}
\bibitem{AyikGregoire}
{S. Ayik and C. Gregoire, Phys. Lett. {\bf B212} (1988)
269; Nucl. Phys. {\bf A513} (1990) 187.}
\bibitem{RandruprRemaud}
{J. Randrup and B. Remaud, Nucl. Phys. {\bf A514} (1990) 339.}
\bibitem{Abe}
Y. Abe, S. Ayik, P.G. Reinhard and E. Suraud, Phys. Rep. {\bf 275} (1996) 49.
\bibitem{Alfio}
{A. Guarnera Thesis, (1996) G.A.N.I.L Caen, {\bf GANIL T.96-01}. \newline
A. Guarnera, M. Colonna and P. Chomaz, Phys. Lett {\bf B373}, (1996) 267.}
\bibitem{Balian}
{R. Balian and M. Veneroni, Ann. Phys. {\bf 135} (1981) 270.}
\bibitem{Suraud}
P.G. Reinhard and E. Suraud, Ann. of Phys. {\bf 216} 
(1992) 98.
\bibitem{Jacquot}
B.  Jacquot, M. Colonna, S. Ayik and Ph. Chomaz, Nucl. Phys. {\bf A617} (1997) 356.
\bibitem{Ayik}
S.  Ayik et al., Z. Phys. {\bf A337} (1990) 413.
\bibitem{Norenberg}
B.  Morgenstern and W. Norenberg, Nucl. Phys. {\bf A492} (1989) 93.
\bibitem{Hofmann}
H. Hofmann and P.J. Siemens, Nucl. Phys. {\bf A257} (1976) 165; {\bf A275} (1977) 464.
\bibitem{AyikZ}
S. Ayik, Z. Phys. {\bf A350} (1994) 45; Adv. in Nuclear Dynamics 2, eds. W. Bauer and G.
Westfall, Plenum Press, New York (1996).
\bibitem{AyikRandrup}
S.  Ayik and J. Randrup, Phys. Rev. {\bf C50} (1994) 2947.
\bibitem{Gardiner}
C.W. Gardiner, Quantum Noise, Springer-Verlag (1991).
\bibitem{Vautherin} {D. Vautherin, J. Treiner and M. Veneroni, 
Phys. Lett. {\bf B191} (1987) 6.}
\bibitem{Lacroix}
D. Lacroix and Ph. Chomaz, preprint GANIL P 98 04 and submitted to
Nucl. Phys. {\bf A} (1998).
\bibitem{Metropolis}
S. E. Koonin, Computational Physics, Addison-Wesley, (1986).
\bibitem{Marie}
N. Marie et al., Phys. Lett. {\bf B391} (1997) 15. \\
M. F. Rivet  et al., Proc. XXXV Int. Winter Meeting in Nuclear Physics, 
ed. I. Iori, Bormio, Italy (1997).
\end{thebibliography}
\end{document}